\documentclass[%
reprint,
superscriptaddress,
 amsmath,amssymb,
 aps,
longbibliography,
floatfix,
]{revtex4-2}

\usepackage[dvipsnames,svgnames,x11names,hyperref]{xcolor}
\usepackage{siunitx}
\usepackage{tabularx}
\usepackage{color}
\usepackage{graphicx}% Include figure files
\usepackage{dcolumn}% Align table columns on decimal point
\usepackage{bm}% bold math
\usepackage{hyperref}% add hypertext capabilities
\usepackage[mathlines]{lineno}% Enable numbering of text and display math
\usepackage{mathrsfs}

\usepackage[centering,hmargin=25mm,tmargin=28mm,bmargin=28mm]{geometry}

\usepackage{braket}

\usepackage[normalem]{ulem}
\hypersetup{colorlinks=true, 
	linkcolor={blue!75!black!80!yellow},
	citecolor={blue!75!black!80!yellow}, 
	urlcolor=black %{blue!75!black!80!yellow}
	}
\makeatletter \renewcommand\@make@capt@title[2]{%
\@ifx@empty\float@link{\@firstofone}{\expandafter\href\expandafter{\float@link}}%
\sffamily{\textbf{#1}}\@caption@fignum@sep#2 }% \makeatother
\usepackage{placeins} % Allows use of the command \FloatBarrier to prevent figures and tables from moving past that point

\begin{document}
\preprint{APS/123-QED}

\title{Hybridized defects in solid-state %low-dimensional 
materials as artificial molecules}

\author{Derek S. Wang}
\thanks{These two authors contributed equally.}
\affiliation{Harvard John A. Paulson School of Engineering and Applied Sciences, Harvard University, Cambridge, MA 02138, USA}

\author{Christopher J. Ciccarino}
\thanks{These two authors contributed equally.}
\affiliation{Harvard John A. Paulson School of Engineering and Applied Sciences, Harvard University, Cambridge, MA 02138, USA}
\affiliation{Department of Chemistry and Chemical Biology, Harvard University, Cambridge, MA 02138, USA}

\author{Johannes Flick}
\affiliation{Center for Computational Quantum Physics, Flatiron Institute, New York, NY 10010, USA}

\author{Prineha Narang}
\email{prineha@seas.harvard.edu}
\affiliation{Harvard John A. Paulson School of Engineering and Applied Sciences, Harvard University, Cambridge, MA 02138, USA}

\begin{abstract}
\noindent Two-dimensional materials can be crafted with structural precision approaching the atomic scale, enabling quantum defects-by-design. These defects are frequently described as ``artificial atoms’' and are emerging optically-addressable spin qubits. However, interactions and coupling of such artificial atoms with each other, in the presence of the lattice, is remarkably underexplored. Here we present the formation of ``artificial molecules’' in solids, introducing a new degree of freedom in control of quantum optoelectronic materials. Specifically, in monolayer hexagonal boron nitride as our model system, we observe configuration- and distance-dependent dissociation curves and hybridization of defect orbitals within the bandgap into bonding and antibonding orbitals, with splitting energies ranging from $\sim$ 10 meV to nearly 1 eV. We calculate the energetics of \textit{cis} and \textit{trans} out-of-plane defect pairs CH$_\textrm{B}$-CH$_\textrm{B}$ against an in-plane defect pair C$_\textrm{B}$-C$_\textrm{B}$ and find that in-plane defect pair interacts more strongly than out-of-plane pairs. We demonstrate an application of this chemical degree of freedom by varying the distance between C$_\textrm{B}$ and V$_\textrm{N}$ of C$_\textrm{B}$V$_\textrm{N}$ and observe changes in the predicted peak absorption wavelength from the visible to the near-infrared spectral band. We envision leveraging this chemical degree of freedom of defect complexes to precisely control and tune defect properties towards engineering robust quantum memories and quantum emitters for quantum information science.
\end{abstract}
\date{\today}

\maketitle

\section{Introduction} \label{sec:intro}
Point defects in both 2D and 3D solids have seen wide applicability in quantum information science~\cite{Wrachtrup2001, Weber2010, Wrachtrup2010, Aharonovich2016, Degen2017, Wang2020}, especially as quantum memories, because they have the potential to combine favorable coherence and non-classical emission properties of isolated atoms~\cite{Kurtsiefer2000, Ye2019} with the scalability and stability of solid-state technologies~\cite{Degen2017, Aharonovich2016, Atature2018, Childress2014}. These defects often introduce localized electronic states that can be effectively decoupled from the host lattice, thereby creating an ``artificial atom''. Such artificial atom systems can include simple substitutional or vacancy defects, as well as more complicated defect complexes~\cite{McDougall2017, MacKoit-Sinkeviciene2019, Czelej2020} that are comprised of multiple imperfections but effectively act as a single defect. To harness the electronic and optical properties of these defects, defects can be coupled to external fields, including electric, magnetic, and also strain, as well as to waveguides and cavity environments~\cite{Rogers2008, Momenzadeh2015, Faraon2012, Chakraborty2019, Zhang2018, Machielse2019, Neuman2020Nanomagnonics, Wang2020Selection}. Despite these advances, demands to the properties of defect systems are ever-increasing in complexity, such as specific level structures for emission of entangled photonic states~\cite{Wang2020, Trivedi2020} or implementation of multi-qubit photonic gates \cite{Dai2020}. Therefore, the need for defect centers with engineered electronic, optical, and spin properties has motivated the search for novel defect centers~\cite{Narang2019, Awschalom2018, Atature2018, Alkauskas2019}. Theory and computationally-led searches have proved particularly valuable here, resulting, for instance, in the discovery of new group IV and prediction of group III emitters in diamond~\cite{Thiering2018, Harris2020, Trusheim2019, Ferrenti2020, Ciccarino2020}.

Discoveries in synthesis and patterning of atomically-architectured materials have so far yielded a diverse portfolio of photonic and optoelectronic quantum materials, including a variety of two-dimensional layered architectures that can be crafted with structural precision approaching the atomic scale, enabling quantum defects-by-design. Inspired by advances in experimental techniques for atom-by-atom manipulation, creation, and observation of individual defects~\cite{Dyck2018, Dyck2018a, Fuechsle2012, Kalinin2019, Ziatdinov2019, Phys2018, Zhang2018, Narang2019, Susi2019, Schuler2019, Schuler2019:2,Tian2020, Scherpelz2017}, we add the crucial chemical degree of freedom to the state-of-the-art in quantum defects by demonstrating how these ``artificial atoms'' become ``artificial molecules'' when placed within a few angstroms of each other and investigate the properties of such coupled defects. We study pairs of substitutional defects within a monolayer of hexagonal boron nitride (hBN), which has a wide bandgap well-suited for hosting well-defined defect orbitals. 

In our calculations, we observe features analogous to molecules, such as the formation of delocalized, bonding and anti-bonding ``molecular'' orbitals arising from the hybridization of the localized, ``atomic" orbitals of the single defects with splitting energies ranging from $\sim 10$ meV to nearly 1 eV. Notably, we determine the energetics of paired in-plane defects and \textit{cis} and \textit{trans} paired out-of-plane defects, showing that the studied paired in-plane defects interact more strongly than out-of-plane ones. With these energies, we generate dissociation curves, and find that they are not merely a function of the inter-defect distance, as is the case in chemical bonding between centrosymmetric atoms, but also the configuration of the defects within the host lattice. Importantly, we find that at these length scales, the host lattice is not only a passive dielectric that screens interactions between defects in complexes, but actively participates in structure relaxation and determines configuration-dependent interactions. 

Armed with this physical understanding, we demonstrate how this chemical degree of freedom can be used to tune quantum emitters. Specifically, we vary the defect-defect distance in C$_\textrm{B}$V$_\textrm{N}$ comprised of C$_\textrm{B}$ and V$_\textrm{N}$, a previously-proposed candidate for emission of visible light from hBN, and observe changes in the predicted peak absorption wavelength from the visible to the near-infrared spectral range. We anticipate that further exploring this chemical degree of freedom of defect complexes with higher order predictive methods will enable new approaches to precisely engineer defect properties as quantum memories and quantum emitters for quantum information science.

\begin{figure}[tbhp]
\centering
\includegraphics[width=0.98\linewidth]{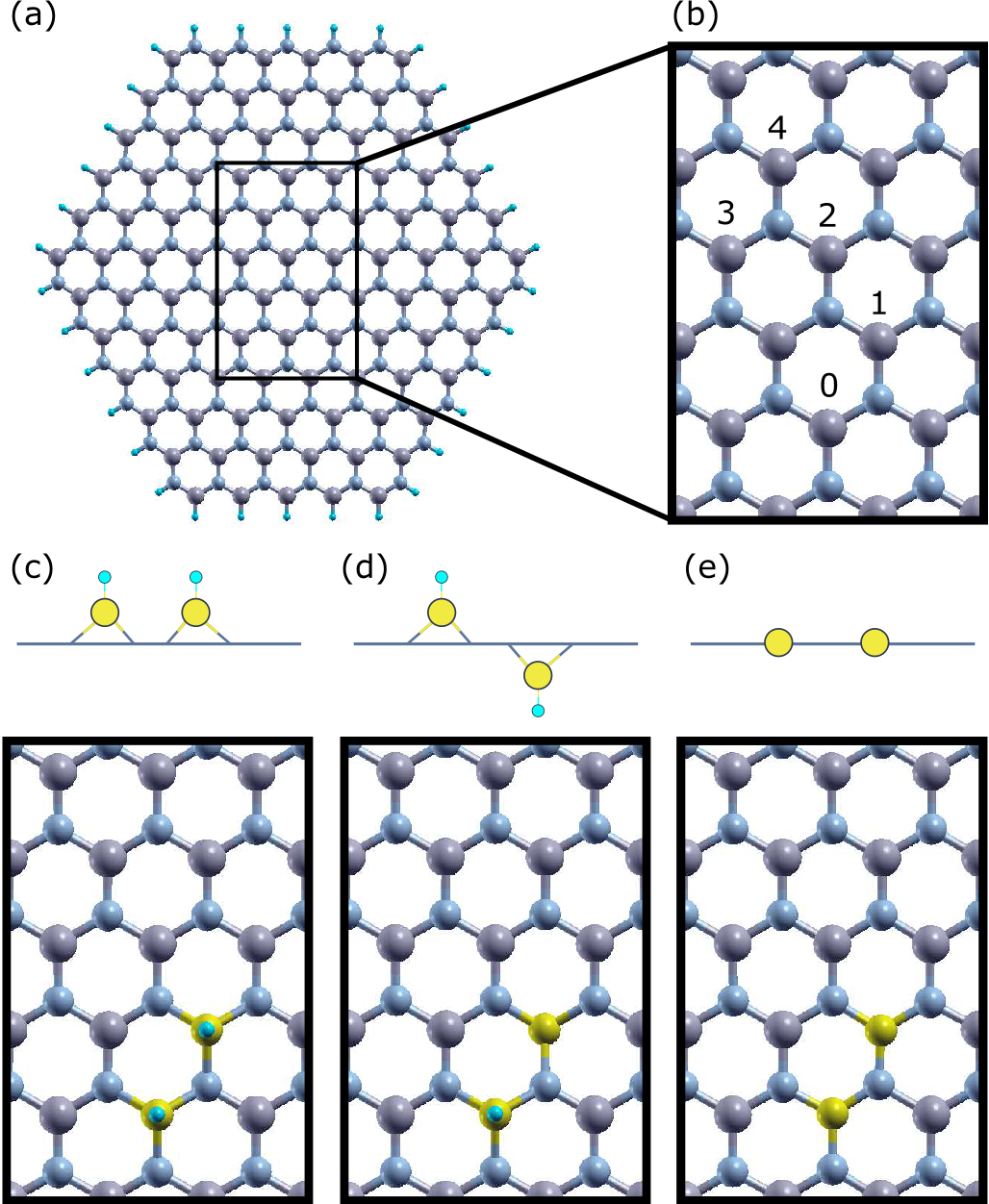}
\caption{Two defects $A_\alpha$-$B_\beta @X$ are placed in \textbf{(a)} a large nanoflake of hBN, where boron is gray and nitrogen is blue. \textbf{(b)} The first defect substituent $A$ replaces the $\alpha$ boron in position 0, and the second defect substituent $B$ replaces the $\beta$ boron atom at position $X$ from 1-4. Out-of-plane and in-plane views of three model double defects \textbf{(c)} \textit{cis} CH$_{\rm B}$-CH$_{\rm B}$@1, \textbf{(d)} \textit{trans} CH$_{\rm B}$-CH$_{\rm B}$@1, and \textbf{(e)} C$_{\rm B}$-C$_{\rm B}$@1, where hydrogen is bright blue and carbon is yellow.
}
\label{fig:schematic}
\end{figure}

\section{Results and Discussion} \label{sec:results}

We model the interactions between two defects spaced up to four unit cells apart in a nanoflake of monolayer hBN large enough to simulate bulk behavior and terminated with hydrogen atoms. For delineating the double defect structure, we adopt the notation to be generally of the form \mbox{$A_\alpha$-$B_\beta @X$}. Here, substituents $A$ and $B$ correspond to the defect substituent or vacancy being introduced. The indices $\alpha$ and $\beta$ represent the atomic sites on which the substituents are placed, which in our case is always boron B for $\alpha$ and either B or nitrogen N for $\beta$. Finally, $X$ corresponds to the location of the second defect marked in the schematic in Fig.~\ref{fig:schematic}(b). To theoretically model these systems, we use the pseudopotential, real-space density functional theory method Octopus~\cite{octopus1, octopus2,octopus3} (see Methods section for computational details). We choose a real-space as opposed to a periodic density functional approach to avoid artificial interactions across neighboring supercells which could obscure the defect-defect interactions of interest. Furthermore, it has been shown that real-space electronic structure calculations on hexagonal boron nitride (hBN) nanoflakes can be extrapolated onto periodic calculations~\cite{Barcza, Reimers2020}. The size of the nanoflake, around 220 atoms in total, is chosen such that there are at least 4 unit cells of hBN beyond the boundaries of the localized defect orbitals before termination with passivating hydrogen atoms. The calculated energy difference between the valence and conduction bands is found to be $E_{\rm bg}=4.28$ eV $\pm 1\%$ for all calculations presented here. 
%Our calculations of electric dipole transition densities of the simulated defect pairs suggest that these defects are not optically active, so in the present study, we consider only the ground state properties of double defect complexes. 

To develop a general understanding of the artificial molecule picture, we first study substituents where $A$ and $B$ are either a carbon atom C or a carbon bonded to a hydrogen CH. Both present a single defect orbital within the band gap, but upon relaxation the $\rm CH_B$ defect adopts an out-of-plane trigonal pyramidal geometry, while the $\rm C_B$ defect adopts an in-plane trigonal planar geometry. From these two single defects $\rm CH_B$ and $\rm C_B$, we compare three distinct pairs of defects. Using the nomenclature defined above, these are \mbox{\textit{cis} CH$_{\rm B}$-CH$_{\rm B}@X$}, \mbox{\textit{trans} CH$_{\rm B}$-CH$_{\rm B}@X$}, and \mbox{C$_{\rm B}$-C$_{\rm B}@X$}. 
Here, $X=1$ to 4 corresponds to lattice sites; see Fig.~\ref{fig:schematic}(a). There is always one defect located at $X = 0$. The corresponding in-plane distances for the defect pairs for $X = 1$ to 4 are 2.62, 4.54, 5.25 and 6.93 \AA, respectively. The out-of-plane and in-plane geometries of these three distinct pairs of defects are shown in Fig.~\ref{fig:schematic}(c)-(e). These defect pairs allow us to generate qualitative understandings of wavefunction overlap in both in- and out-of-plane molecular configurations. % would be cool to say "molecular" configurations.

\begin{figure}[tbhp]
\centering
\includegraphics[width=0.98\linewidth]{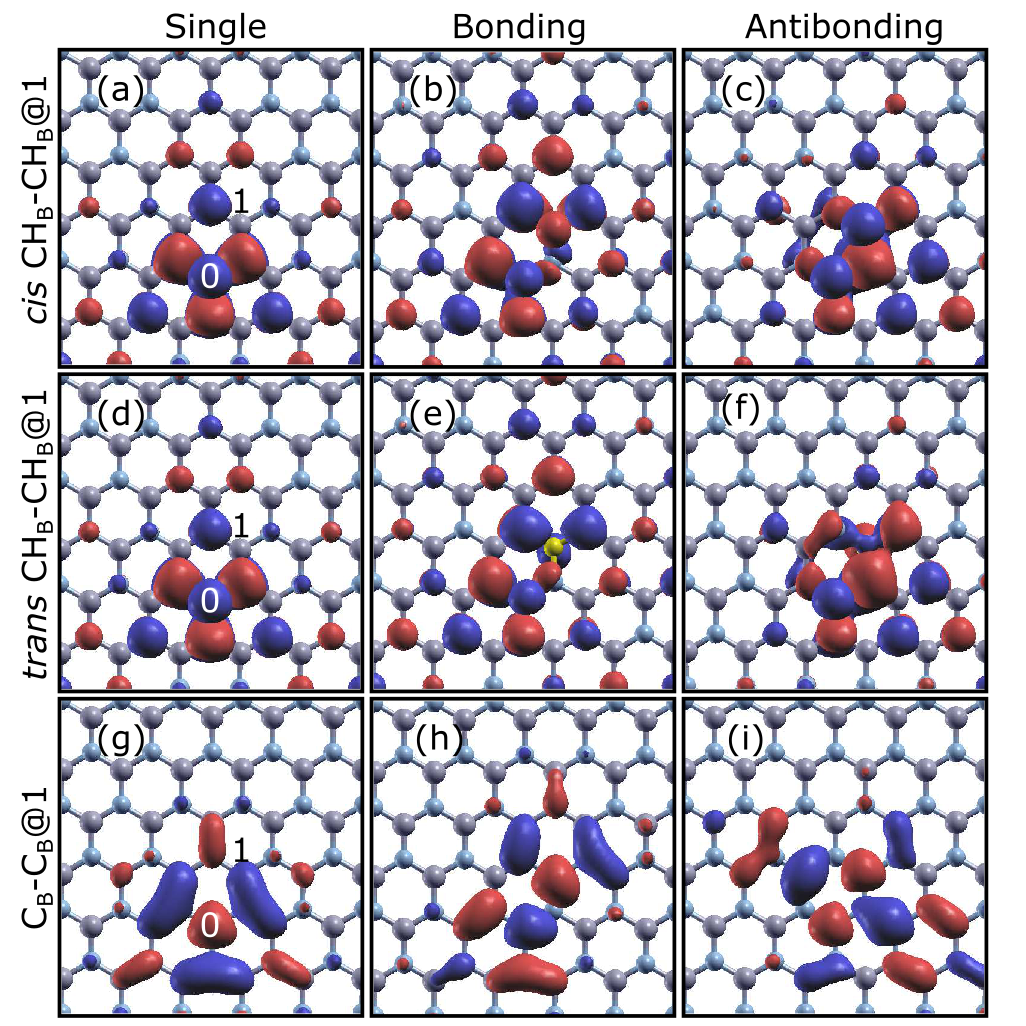}
\caption{Orbital of single defects within the band gap for \textbf{(a)} and \textbf{(d)} CH$_{\rm B}$ and \textbf{(g)} C$_{\rm B}$. Bonding (middle) and anti-bonding (right) defect orbitals, corresponding to lower and higher energy, within the bandgap for the nearest defect-defect positions calculated for \textbf{(b)-(c)} \mbox{\textit{cis} CH$_{\rm B}$-CH$_{\rm B}@1$}, \textbf{(e)-(f)} \textit{trans} CH$_{\rm B}$-CH$_{\rm B}@1$, and \textbf{(h)-(i)} C$_{\rm B}$-C$_{\rm B}@1$. Red and purple colors of the wavefunction isosurfaces correspond to positive and negative values, respectively.}
\label{fig:orbitals}
\end{figure}

The single defects CH$_{\rm B}$ and C$_{\rm B}$ discussed above each contribute one orbital in the band gap between the highest occupied (``valence band'') and lowest unoccupied delocalized Kohn-Sham states (``conduction band''), plotted in Fig.~\ref{fig:orbitals}(a)/(d) and Fig.~\ref{fig:orbitals}(g), respectively. When a second defect is added, we find two orbitals collectively that are separated in energy from each other. In the spirit of molecular orbital theory, we denote the lower energy one the ``bonding orbital'' and the higher one the ``antibonding orbital.'' Figure~\ref{fig:orbitals} compares the orbital character of these defect orbitals of the most strongly interacting configuration of each defect pair, where $X=1$. For all three defect pairs, both the bonding and antibonding orbitals are delocalized over both defect sites, supporting the ``artificial molecules'' picture where the defects have hybridized. 
%We also note that the bonding and antibonding orbitals of $\rm C_B$-$\rm C_B$ (Fig.~\ref{fig:orbitals}(e) and (f)) are clearly oppositely polarized in space, suggesting that optical transitions to the bulk bands from the defect orbitals in the single particle picture will be oppositely polarized.

\begin{figure}[tbhp]
\centering
\includegraphics[width=0.98\linewidth]{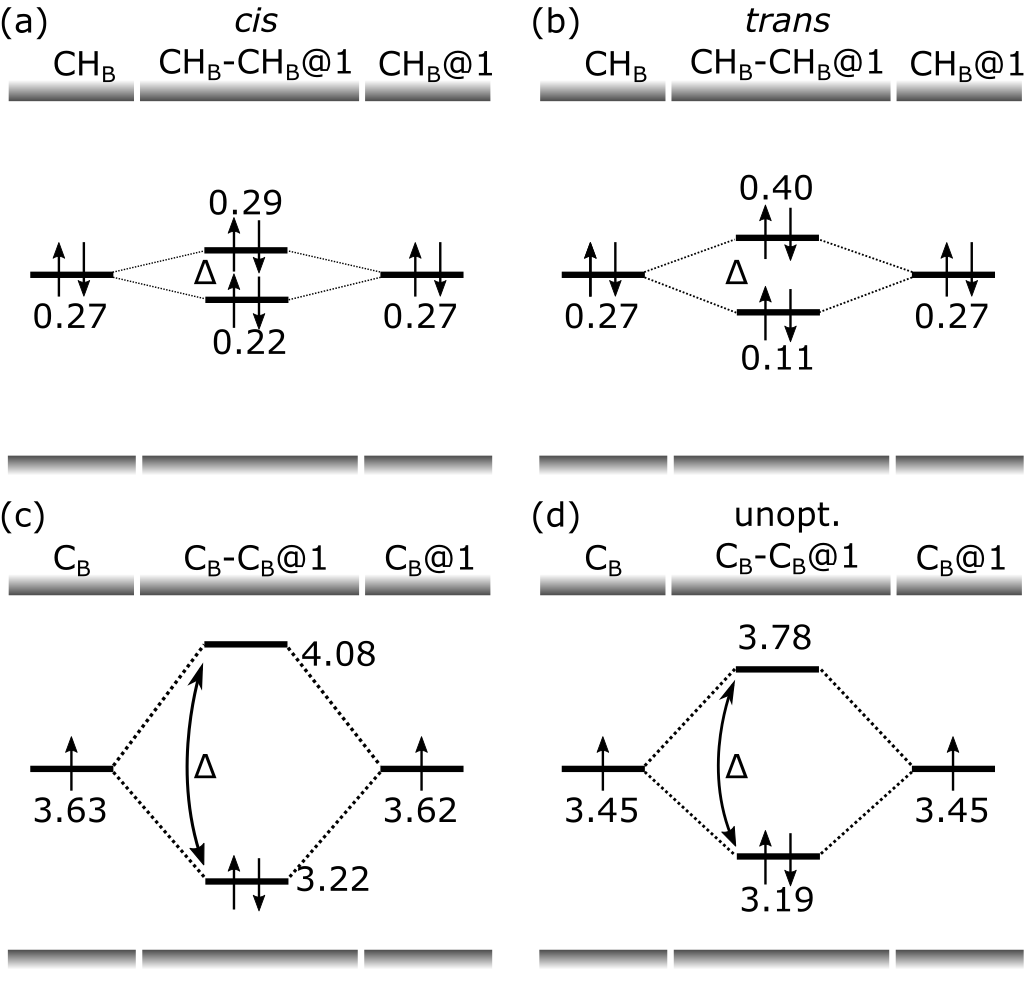}
\caption{Energy, hybridization, and occupation of single defect orbitals and hybridized double defect orbitals for \textbf{(a)} \textit{cis} CH$_{\rm B}$-CH$_{\rm B}@1$, \textbf{(b)} \textit{trans} CH$_{\rm B}$-CH$_{\rm B}@1$, \textbf{(c)} C$_{\rm B}$-C$_{\rm B}@1$, and \textbf{(d)} unoptimized C$_{\rm B}$-C$_{\rm B}@1$. The splitting energy is $\Delta$. The bottom and top gray bars correspond to the valence and conduction bands, respectively. All values shown are in eV.
}
\label{fig:MO}
\end{figure}

We further understand the implications of these hybridized defect orbitals on the electronic structure by analyzing the orbital splitting energies and occupations of the single defects vs. the paired defects in molecular orbital-like diagrams at $X=1$ in Fig.~\ref{fig:MO}. The energies and occupations of the orbitals of the single defects CH$_{\rm B}$ and C$_{\rm B}$ within the bandgap are plotted on the left and right in Fig.~\ref{fig:MO}(a)/(b) and Fig.~\ref{fig:MO}(c), respectively. Their energetic positions are in qualitative agreement with results in Ref.~\citenum{McDougall2017}. Upon hybridization of two single CH$_{\rm B}$ or C$_{\rm B}$ defect orbitals in the double defects, at $X=1$, we see that the double defect orbitals are symmetrically split by $\Delta=70$, 290, and 860 meV around the energy of the single defect orbital for \textit{cis} $\rm CH_B$-CH${\rm_B}@1$, \textit{trans} $\rm CH_B$-CH${\rm_B}@1$, and $\rm C_B$-C${\rm_B}@1$ in the middle of Fig. \ref{fig:MO}(a), (b), and (c), respectively. The splitting energy $\Delta$ of the in-plane paired defects $\rm C_B$-C${\rm_B}@1$ is several times higher than the out-of-plane defect pairs, \textit{cis} and \textit{trans} $\rm CH_B$-C${\rm_B}@1$. The occupation of these artificial molecular orbitals is based on the total number of electrons contributed by the individual defects, where the single defect orbital of $\rm CH_B$ or $\rm C_B$ is doubly or singly occupied for the charge neutral configuration in the absence of spin-polarization. The paired defects then result in full occupation of the artificial molecular orbitals of \textit{cis} and \textit{trans} CH${\rm_B}$-CH${\rm_B}@1$ and occupation of only the bonding orbital of C${\rm_B}$-C${\rm_B}@1$.  

The hybridization and occupation of the defect orbitals can be analogized to the canonical examples of bonding between two hydrogen or helium atoms to form H$_2$ or He$_2$, where H$_2$ forms a covalent bond, while He$_2$ interacts via weaker dispersion interactions. In these cases, orbital hybridization similarly results in a lower energy bonding orbital and a higher energy antibonding orbital. In the gas phase, only H$_2$ is stable, however, due to the full occupation of only the lower energy bonding orbital. We can therefore anticipate the $\rm C_B$-C${\rm_B}@X$ defect pair to become more stable upon defect orbital hybridization. Furthermore, as the atom-atom distance increases beyond the equilibrium bond position, we can also expect the potential energy of the the $\rm C_B$-C${\rm_B}@X$ defect pair to increase as the orbital overlap decreases.

\begin{figure*}[tbhp]
\centering
\includegraphics[width=0.98\linewidth]{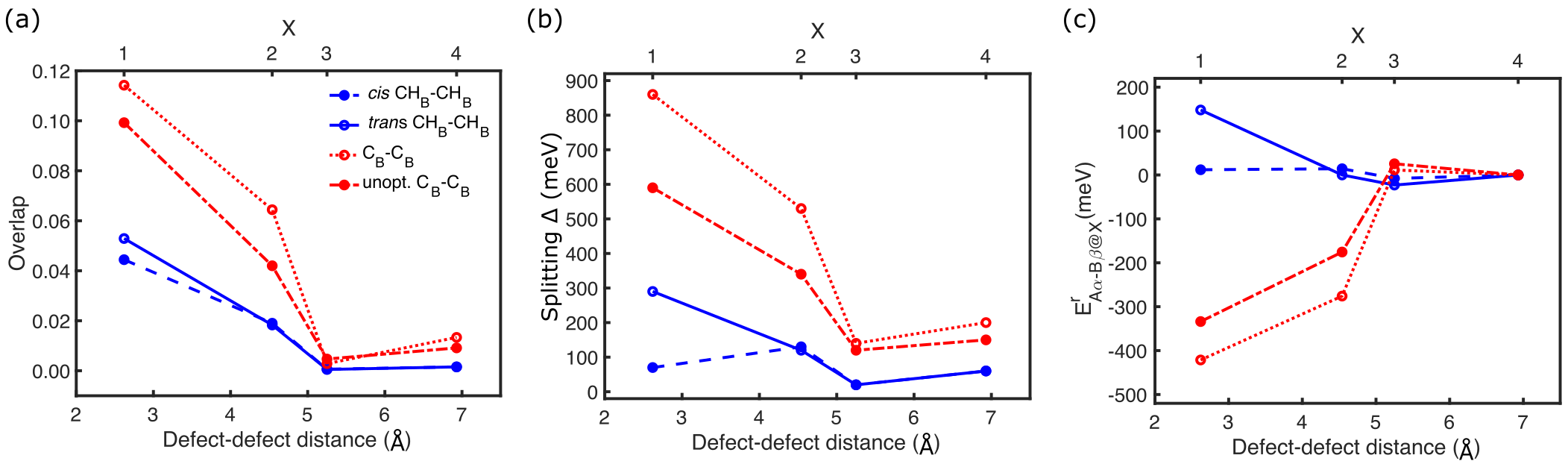}
\caption{\textbf{(a)} Wavefunction overlaps, \textbf{(b)} splittings $\Delta$ of bonding-antibonding orbitals, and \textbf{(c)} relative energies of formation $E^r_{A_\alpha\textnormal{-}B_\beta @X}$ for \textit{cis} CH$_{\rm B}$-CH$_{\rm B}@X$, \textit{trans} CH$_{\rm B}$-CH$_{\rm B}@X$, C$_{\rm B}$-C$_{\rm B}@X$, and unoptimized C$_{\rm B}$-C$_{\rm B}@X$.
}
\label{fig:distance}
\end{figure*}

We test this analogy by calculating important signatures of orbital hybridization for increasing defect-defect distances in Fig.~\ref{fig:distance}. In Fig.~\ref{fig:distance}(a), we plot the wavefunction overlaps, calculated as the modulus squared of the inner product of single defect orbitals placed at $X=0$ and $X=1$ to 4. Based on the molecular bonding analogy, since we observe larger $\Delta$ for the in-plane defect pair C${\rm_B}$-C${\rm_B}@1$ than for the out-of-plane defect pairs \textit{cis} and \textit{trans} CH${\rm_B}$-CH${\rm_B}@X$, we should observe larger overlaps for the in-plane paired defects than the out-of-plane ones, and indeed we do for all $X$. Based further on the molecular analogy of bonding centrosymmetric atoms, we should also observe lower overlap with increasing defect-defect distance. From $X=1$ to 3 where the defect-defect distances increases from 2.62 to 5.25 \AA, as expected, the overlap decreases. However, at $X=4$, we see for all defect pairs that the overlap does not continue to decrease, but rather increases. This trend reversal can be attributed to the non-centrosymmetric nature of the defect orbitals. Evidently the $X=4$ position is more conducive than $X=3$ for overlap between the lobes of the three-fold symmetric defect orbitals, despite the larger defect-defect distance. The trends in splitting energies $\Delta$ indicating the degree or orbital hybridization generally mirror those of the overlaps: the in-plane defect pairs have larger overlaps and thus larger $\Delta$ than the out-of-plane ones, and $\Delta$ decreases with increasing defect-defect distances until $X=3$ to $X=4$, where both the overlap and splitting are higher for the latter. The only notable exception is \textit{cis} CH${\rm_B}$-CH${\rm_B}@X$, where the overlap increases but splitting decreases from from $X=1$ to $X=2$. Recalling that overlap is calculated for single defects at positions 0 and $X=1$ to 4, steric effects may prevent effective overlap in this paired defect, resulting in lower $\Delta$ at $X=1$.

In Fig.~\ref{fig:distance}(c), we plot dissociation curves for the paired defects, computing the relative energy of formation $E_{A_\alpha\textnormal{-}B_\beta @X}^{\rm r}$ relative to the energy $E_{A_\alpha\textnormal{-}B_\beta @X}$ of the geometry corresponding to the farthest defect-defect distance at $X=4$. We also take into account systematic energy differences for single defects placed at positions $X=1$ to 4. The overall expression for $E_{A_\alpha\textnormal{-}B_\beta @X}^{\rm r}$ is:
\begin{align}
    E_{A_\alpha\textnormal{-}B_\beta @X}^{\rm r} &= (E_{A_\alpha\textnormal{-}B_\beta @X}-E_{A_\alpha\textnormal{-}B_\beta @4}) \nonumber
    \\ &+(E_{B_\beta @X}-E_{B_\beta @4}).
\end{align}

Based on the molecular analogy, for the $\rm CH_B$-CH${\rm_B}@X$ defects with fully occupied bonding and antibonding orbitals symmetrically split in energy around the energy of the single defect orbitals, barring any other effects, we would expect low-magnitude energies of formation $E^r_{A_\alpha\textnormal{-}B_\beta @X}$. Indeed, the energy of formation of \mbox{\textit{cis} $\rm CH_B$-CH${\rm_B}@X$} is barely impacted by the defect-defect interactions. However, its \textit{trans} counterpart in Fig.~\ref{fig:distance}(b) experiences a substantial energy barrier of $\sim$ 140 meV for the nearest possible interaction at $X=1$. Beyond $X=1$, the defect-defect interaction quickly drops off. The large formation energy of \textit{trans} $\rm CH_B$-$\rm H_B@1$ may therefore be due to a locally strained environment, demonstrating the importance of the host lattice in determining energetics. These trends for the out-of-plane defects are in sharp contrast to that of $\rm C_B$-C${\rm_B}@X$ shown in Fig.~\ref{fig:distance}(c), where the energy of formation is opposite in sign and three times larger in magnitude at $X=1$ compared to the energy of formation of \textit{trans} $\rm CH_B$-CH${\rm_B}@X$, suggesting that $\rm C_B$ defects are more stable when paired closely together because the lower energy bonding orbital is occupied while the higher energy antibonding orbital is left unoccupied. We also note that sign of $E^r_{A_\alpha\textnormal{-}B_\beta @X}$ flips from $X=2$ to $X=3$ before returning to 0 at $X=4$ by definition, whereas in a traditional molecular dissociation curve, the potential simply approaches 0 with no sign flip. We can attribute this feature to the same trend reversal in overlap and $\Delta$ discussed previously, overall illustrating that overlap, $\Delta$, and $E^r_{A_\alpha\textnormal{-}B_\beta @X}$ depend not only on the defect-defect distance but also the geometry. We note that we also investigate the impact of spin-polarization on the open-shell C$_{\rm B}$ single defect and determine that the SCF energy is uniformly lowered by $\sim 270$ meV for C$_{\rm B}$@X where $X=1$ to 4, resulting in no change to the relative formation energy of the double defects $E_{A_\alpha-B_\beta @X}^r$.

While the analogy of defects hybridizing as atoms do to form molecules can be convenient, the host lattice can non-trivially impact the electronic structure of the defects. To determine the quantitative effect of host lattice relaxation, we show the orbital energies, occupations, and relative formation energies for the paired defect C$_{\rm B}$-C$_{\rm B}$@$X$ and the single defect $\rm C_B$ in Fig.~\ref{fig:MO}(d) and Fig.~\ref{fig:distance}(d) where the C$_{\rm B}$ defects are inserted into a relaxed, pristine nanoflake of hBN and not geometry optimized again afterward. Similar to its optimized counterpart, the single defect orbital is singly occupied, but its energy relative to the valence band is shifted downward by 180 meV. The orbitals of the double defect are also split generally evenly around this energy, as in the three, optimized model double defects. Notably, however, the splitting is in general smaller than in the relaxed C$_{\rm B}$-C$_{\rm B}$@$X$ double defects, suggesting that host lattice relaxation enables more efficient overlap of the single defect orbitals. The relative decrease in splitting then causes the magnitude of the relative formation energy to be smaller for close defect-defect distances at $X=1$ and $X=2$. These results highlight the importance of including both the defect complex and host lattice in computational studies and not simply computing the hybridization of the orbitals of single defects to predict properties of defect complexes.

\begin{figure*}[tbhp]
\centering
\includegraphics[width=0.98\linewidth]{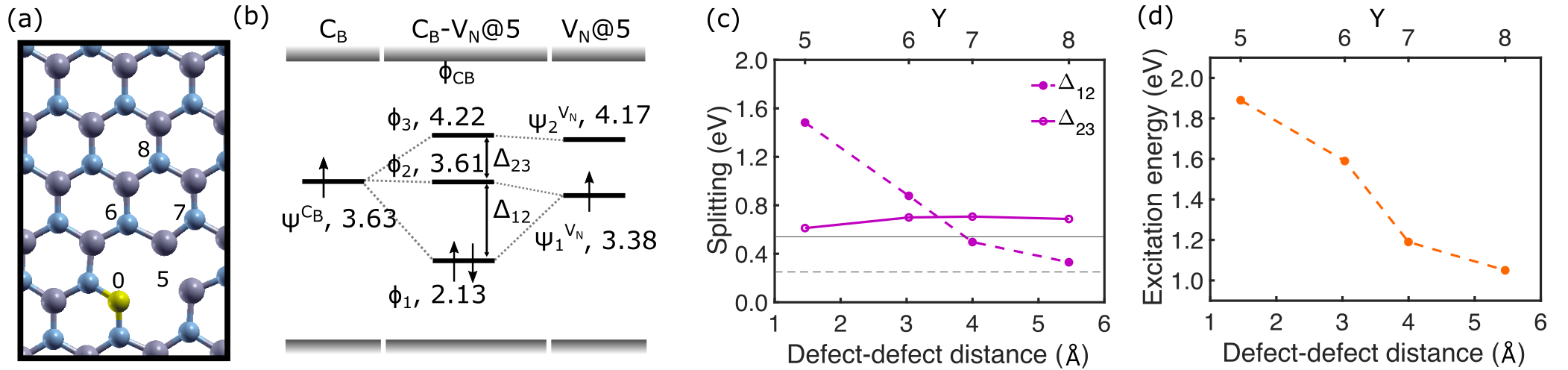}
\caption{Double defect and emitter C$_\textrm{B}$-V$_\mathrm{N}@Y$. \textbf{(a)} Geometry of the double defect C$_\textrm{B}$-V$_\mathrm{N}@5$, or C$_\textrm{B}$V$_\mathrm{N}$, in the same sized nanoflake of hBN as shown in Fig.~\ref{fig:schematic}. For all double defects C$_\textrm{B}$-V$_\mathrm{N}@Y$, C$_\textrm{B}$ is at position 0 and V$_\textrm{N}$ is at position $Y$ from 5 to 8. Boron is gray, nitrogen is blue, and carbon is yellow. \textbf{(b)} Molecular orbital diagram of C$_\textrm{B}$-V$_\mathrm{N}@5$ and its single defect constituents C$_\textrm{B}$ and V$_\mathrm{N}$, where all localized Kohn-Sham orbitals are labelled, as well as the lowest-lying level in the ``conduction" band $\phi_\mathrm{CB}$. \textbf{(c)} Splittings $\Delta_{12}$ and $\Delta_{23}$ between the defect orbitals of C$_\textrm{B}$-V$_\mathrm{N}@Y$. The lower, dashed and upper, solid black lines correspond to energy differences between $\psi^{\mathrm{V_N}}_1\leftrightarrow\psi^{\mathrm{C_B}}$ and $\psi^{\mathrm{C_B}}\leftrightarrow\psi^{\mathrm{V_N}}_2$, respectively, in the limit of zero hybridization. \textbf{(d)} Excitation energy of the lowest-lying excitation in C$_\textrm{B}$-V$_\mathrm{N}@Y$ for $Y=5$ to 8.
%and colored according to dominant transition densities.
}
\label{fig:emitter}
\end{figure*}

We leverage our general understanding of the picture of bonding artificial atoms into artificial molecules to design new quantum emitters with the chemical degree of freedom. As an example, we study derivatives of the neutral C$_\textrm{B}$V$_\mathrm{N}$ defect. This well-studied defect in hBN is hypothesized to be a bright emitter of visible light~\cite{Tawfik2017, Abdi2018} with potential applications as a single photon emitter in technologies. C$_\textrm{B}$V$_\mathrm{N}$ is conventionally treated as a single defect, but here, we consider it a defect complex C$_\textrm{B}$-V$_\mathrm{N}@Y$ of single defects C$_\textrm{B}$ and V$_\mathrm{N}$, where C$_\textrm{B}$ is at $X=0$ and V$_\mathrm{N}$ is placed at $Y=5$ to 8, as labelled in Fig.~\ref{fig:emitter}(a). These geometries correspond to in-plane distances ranging from 1.46 \AA~to 5.46 \AA. In our notation, therefore, C$_\textrm{B}$V$_\mathrm{N}$ is equivalent to C$_\textrm{B}$-V$_\mathrm{N}@5$. As shown in Fig.~\ref{fig:emitter}(b), C$_\textrm{B}$-V$_\mathrm{N}@Y$ has three defect orbitals within the band gap, labelled $\phi_1$, $\phi_2$, and $\phi_3$ in order of increasing energy. One is from C$_\textrm{B}$ ($\psi^{\mathrm{C_B}}$) and two are from V$_\mathrm{N}@Y$ ($\psi^{\mathrm{V_N}}_1$ and $\psi^{\mathrm{V_N}}_2$). We also label the lowest-lying conduction orbital $\phi_\mathrm{CB}$. Because $\phi_\mathrm{CB}$ and $\phi_3$ lie so closely in energy (a few meV apart), numerical error results in some mixing in these two orbitals between the localized character of a defect orbital and the delocalized character of a bulk-like band state. The single defect constituents C$_\textrm{B}$ and V$_\mathrm{N}$ each contribute one electron to the double defect C$_\textrm{B}$-V$_\mathrm{N}@Y$, resulting in full occupation of the lowest energy defect orbital. As with the previously discussed in- and out-of-plane defect pairs, we find that adjusting the relative geometry from $Y=5$ to 8 of the single defects C$_\mathrm{B}$ and V$_\mathrm{N}$ results in energy splitting $\Delta_{12}$ between the localized defect orbitals of C$_\mathrm{B}$-V$_\mathrm{N}@Y$, up to 1.48 eV for $Y=5$ as plotted in Fig.~\ref{fig:emitter}(c) and closely matching the results of Ref.~\citenum{Tawfik2017}. Notably, however, $\Delta_{23}$ is largely unchanged by varying $Y$, as $\psi^\mathrm{C_B}$ and $\psi_2^\mathrm{V_N}$ are well-separated in energy, resulting in less splitting for a given perturbation. Finally, as expected, as the defects move farther apart and hybridize less, $\Delta_{12}$ and $\Delta_{23}$ approach the energy differences between $\psi^{\mathrm{V_N}}_1\leftrightarrow\psi^{\mathrm{C_B}}$ and $\psi^{\mathrm{C_B}}\leftrightarrow\psi^{\mathrm{V_N}}_2$, respectively, in the limit of zero hybridization as the lower, dashed and upper, solid black lines in Fig.~\ref{fig:emitter}(c), which correspond to the energy differences between the single defect orbitals in Fig.~\ref{fig:emitter}(b).
%As with the other double defect pairs discussed previously and as shown in Fig. \ref{fig:emitterdistance} in Appendix \ref{app:CBVN}, the degree of spitting $\Delta_{12}$ and $\Delta_{23}$ is strongly dependent on wave function overlap of the single defect orbital constituents, which leads to a lower energy of formation for the closest defect-defect distance at $Y=5$. Note that the splitting $\Delta_{12}$ is generally larger than $\Delta_{23}$ because the sole defect orbital of the former with energy of 3.63 eV above the valence band can more strongly interact with the defect orbital that lies closer in energy, the lower energy defect orbital of the latter with energy of 3.38 eV, than the higher energy defect orbital with energy 4.17 eV, as expected from conventional perturbation theory.

To demonstrate the impact of bonding on the optical activity of defects, in Fig.~\ref{fig:emitter}(d), we plot the excitation energies for the lowest-lying excitation of C$_\textrm{B}$-V$_\textrm{N}@Y$ for $Y=5$ to 8 calculated within the linear-response framework (Casida method) of the time-dependent density functional theory~\cite{casida1995time, flick2019lmrnqe, flickexcited, Wang2020LossQEDFT} and discussed further in Appendix~\ref{app:compmethods}. From $Y=5$ to $Y=8$, the energy of this transition follows the same trend in splitting $\Delta_{12}$, decreasing from 1.89 eV to 1.19 eV, or from red of the visible spectrum to the near-infrared. These results indicate that angstrom-level control of defect geometry may be a powerful method for tuning the excitation energies of defect emitters. 
%With the Casida method, we can also determine which orbital-orbital transitions dominate the optical activity of each excitation. For the lowest-lying excitation described, we find that at $Y=5$, where C$_\mathrm{B}$ and V$_\mathrm{N}$ interact strongly, the absorption intensity is dominated in character by $\phi_1\leftrightarrow\phi_2$, or a transition between strongly hybridized orbitals with a previously noted large splitting $\Delta_{12}$ of 1.48 eV. At $Y=8$, where C$_\mathrm{B}$ and V$_\mathrm{N}$ interact the least strongly of the geometries studied in this work, the absorption intensity is instead dominated by the $\phi_1\leftrightarrow \phi_\mathrm{VB}$, suggesting little excited-state interaction between the single defect constituents. Finally, we note that $Y=6$ exhibits an absorption intensity nearly an order of magnitude higher than the other geometries $Y=5, 7$ and 8 with approximately equal contributions from transitions between $\phi_1$ and the three lowest-lying empty Kohn-Sham orbitals $\phi_2$, $\phi_3$, and $\phi_\mathrm{VB}$, highlighting the potentially drastic impacts of the chemical degree of freedom on the optical activity of defect centers.

\section{Conclusions and Outlook}
The presented results support the picture of artificial molecules formed by bonding artificial atoms comprised of defects with localized orbitals within the band gap of a low-dimensional semiconductor or insulator. We theoretically describe specific defect pairs with varying defect-defect distances in monolayer flakes of hexagonal boron nitride that remain either in-plane or move out-of-plane. We show that orbital overlap-induced orbital splitting ranges from the order of tens of millielectronvolts to nearly 1 eV, realized via a pair of in-plane defects, $\rm C_B$-$\rm C_B$, at the closest defect-defect distance $X=1$ corresponding to $2.62$ \AA. In this particular case, due to full occupation of the lower energy bonding orbital and empty occupation of the higher energy antibonding orbital, the bonded defect pair is more energetically favorable than when the defects are separated farther apart and interact weakly by several hundreds of millielectronvolts. In essence, we observe an effective chemical bond between defect orbitals. These concepts can be used to design novel quantum emitters with custom optical properties. As a example, we show that the absorption wavelength of C$_\textrm{B}$-V$_\mathrm{N}@Y$ can be tuned from the visible to the near-infrared range at different relative defect-defect geometries $Y$.

The results of this study drive two major conclusions. The first is that this study further illustrates how the formation of artificial molecules, or multi-defect complexes, could be responsible for unassigned zero phonon lines in the emission of defect centers in hBN~\cite{Bommer2019}, and the observation that dramatic shifts in the emission line of a given defect center cannot be explained with local strain alone~\cite{Hayee2020}. The second conclusion is that this chemical degree of freedom, coupled with recent experimental innovations that enable deterministic placement of defect centers, offers a novel approach for precisely engineering the optical properties of solid-state quantum emitters, properties that cannot be readily accessed via coupling to cavities for example.

We anticipate that our predictions will motivate experimental studies towards systematic synthesis and characterization of artificial molecules in solid-state materials via techniques with atomic precision. Promising directions for further theoretical studies include inclusion of electron-phonon coupling~\cite{Norambuena2016,Alkauskas2014,Jara2020}, charge state dynamics and charge transition levels~\cite{Freysoldt2014,Dreyer2018,Harris2020}, Jahn-Teller effects~\cite{Ciccarino2020,Thiering2018,Thiering2019,Ivady2020} and application of group theoretical methods~\cite{Abdi2018}, especially to describe the properties of optically excited states. We expect these results to spur further fundamental and technological advances in the use of quantum defects-by-design in solid-state materials for applications in quantum technologies.

\appendix
\section{Computational methods} \label{app:compmethods}

To theoretically model these systems, we use a pseudopotential, real-space density functional theory (DFT) code Octopus~\cite{octopus1, octopus2,octopus3}.
Optimized molecular geometries and ground state electronic densities are calculated with SG15 optimized norm-conserving Vanderbilt pseudopotentials~\cite{Schlipf2015, Hamann2013} and the Perdew, Burke, and Ernzerhof (PBE) generalized gradient approximation exchange-correlation functional~\cite{Perdew1996}, which has been used in previous studies on the electronic and optical properties of defects in hBN~\cite{McDougall2017, Tawfik2017, Tran2016a, Wu2017, Tancogne-dejean2018}. Note that while the PBE functional is known to underestimate the bandgap and systematically mischaracterize certain optical properties~\cite{Abdi2018}, 
results from PBE often qualitatively match those from the more accurate and computationally expensive Heyd–Scuseria–Ernzerhof functional (HSE) functional~\cite{Ernzerhof2003}. The real-space simulation box is parameterized with a mesh spacing of 0.20~\AA, as in Ref.~\citenum{Tancogne-dejean2018}, and consists of spheres with 4~\AA~radius around each atom. The ground state self-consistent field (SCF) energies are converged below 1 meV/atom, and the energy difference $E_\mathrm{bg}$ between the highest occupied and lowest unoccupied bulk-like Kohn-Sham orbitals is converged within 1\% of its mean value of 4.28 eV for nanoflakes with either a single or a pair of defects. $E_\mathrm{bg}$ agrees with the range of bandgaps calculated for nanoflakes of hBN with approximately 100 boron and nitrogen atoms with either nitrogen or boron edges~\cite{Maruyama2018}, in contrast to a calculated bandgap of 4.50 eV with periodic DFT codes and the PBE functional~\cite{Tancogne-dejean2018}. 

\begin{figure}[tbhp]
\centering
\includegraphics[width=0.98\linewidth]{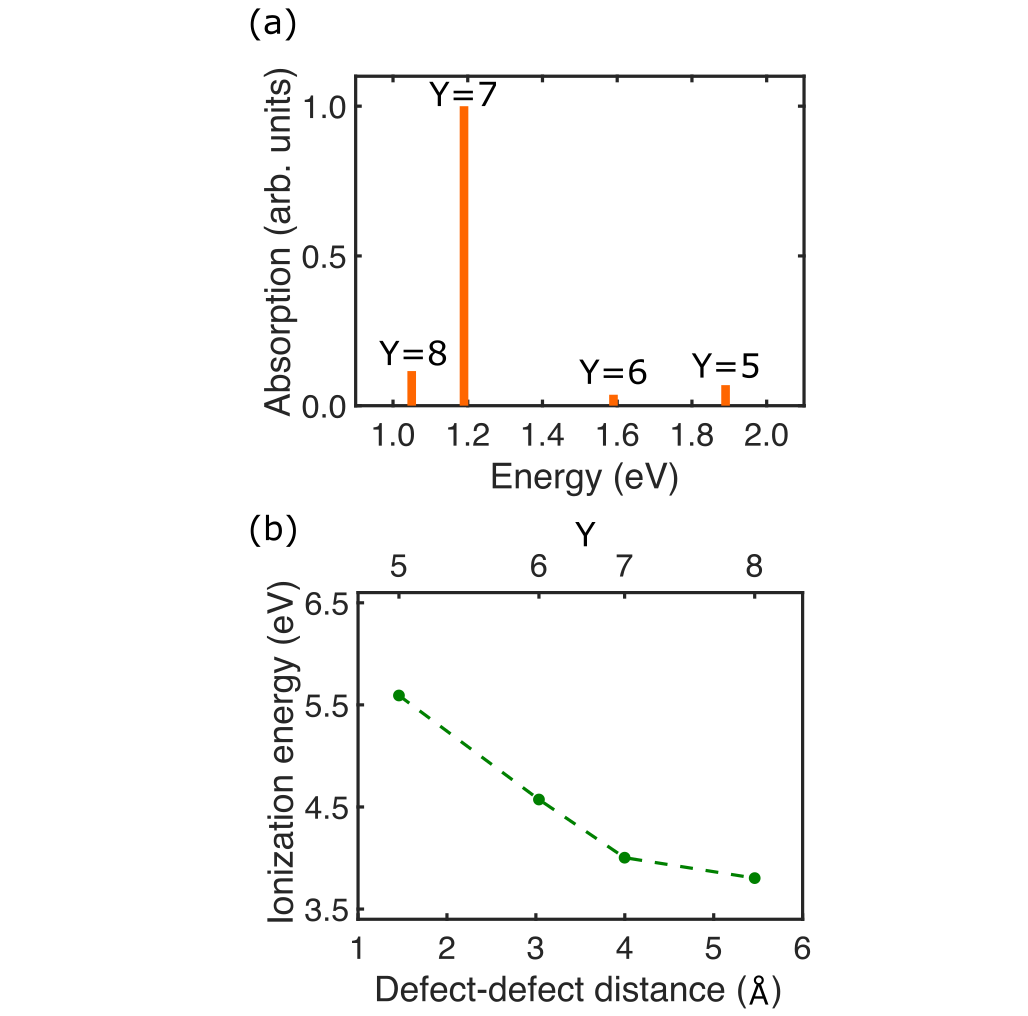}
\caption{\textbf{(a)} Absorption spectra and \textbf{(b)} ionization energies of the double defect C$_\textrm{B}$-V$_\mathrm{N}@Y$ from $Y=5$ to 8.
}
\label{fig:emitterappendix}
\end{figure}

The energies of the excited states of interest converged with 100 extra states in the Casida method, a standard linear-response time-dependent density-functional theory (TDDFT)~\cite{casida1995time}. The resulting absorption spectrum of the lowest-lying excitation is plotted in Fig.~\ref{fig:emitterappendix}(a), where the energies correspond to those plotted in Fig.~\ref{fig:emitter}(d) and the absorption is proportional to the squared transition dipole moment times the transition frequency~\cite{flick2019lmrnqe, flickexcited, Wang2020LossQEDFT}. We confirm that the adiabatic LDA functional is a valid approximation within the Casida formalism by checking that the ionization energies plotted in Fig.~\ref{fig:emitterappendix}(b) and determined with a $\Delta$SCF calculation are several eV higher than the excitation energies in the absorption spectra. Notably, we find that the relative absorption intensities for varying $Y$ depend on the level of linear-response time-dependent density functional theory applied. Specifically, the Petersilka~\cite{Petersilka1996} and Tamm-Dancoff~\cite{Hirata1999} approximations to the full Casida method result in different absorption orderings, a well-known issue~\cite{Hu2014}. Resolving these discrepancies motivates the development of a group theoretical approach to predict the optical activity of artificial molecules whose symmetry strongly depends on the host lattice, defect-defect interactions, and in-plane/out-of-plane movements. Despite these differences in absorption intensities, all three methods predict similar transition energy orderings. Structures and Kohn-Sham orbitals are visualized with XCrySDen~\cite{Kokalj1999}. 

%\section{C$_\textrm{B}$-V$_\textrm{N}$@Y} \label{app:CBVN}

%\begin{figure*}[tbhp]
%\centering
%\includegraphics[width=0.98\linewidth]{figure_emitter_distance.png}
%\caption{Caption.
%}
%\label{fig:emitterdistance}
%\end{figure*}

\section*{Acknowledgements}
D.S.W. and C.J.C contributed equally to this work. We acknowledge fruitful discussions with Tom\'a\v{s} Neuman, John Philbin, and Chitraleema Chakraborty. This work was supported by the Department of Energy `Photonics at Thermodynamic Limits’ Energy Frontier Research Center under grant DE-SC0019140 (theoretical and computational approaches in light-matter interactions), by the U.S. Department of Energy, Office of Science, Basic Energy Sciences (BES), Materials Sciences and Engineering Division under FWP ERKCK47 `Understanding and Controlling Entangled and Correlated Quantum States in Confined Solid-state
Systems Created via Atomic Scale Manipulation' (spin qubits), and by the the Army Research Office MURI (Ab Initio Solid-State Quantum Materials) grant number W911NF-18-1-0431 (computational methods for low dimensional materials). D.S.W. is an NSF Graduate Research Fellow. P.N. is a Moore Inventor Fellow through Grant GBMF8048 from the Gordon and Betty Moore Foundation. The Flatiron Institute is a division of the Simons Foundation.
This research used resources of the National Energy Research Scientific Computing Center, a DOE Office of Science User Facility supported by the Office of Science of the U.S. Department of Energy under Contract No.DE-AC02-05CH11231. Additional calculations were performed using resources from the Department of Defense High Performance Computing Modernization program as well as resources at the Research Computing Group at Harvard University. 

\newcommand{\noopsort}[1]{} \newcommand{\printfirst}[2]{#1}
  \newcommand{\singleletter}[1]{#1} \newcommand{\switchargs}[2]{#2#1}

\end{document}